\documentclass[aps,prl,twocolumn,showpacs,supergroupedaddress,epsfig,amsmath,amssymb]{revtex4}
\usepackage{epsfig,amsmath,amssymb,bm,epsf,graphics,psfrag,verbatim,subfigure}
\usepackage{bbm}
\usepackage{mathrsfs}
\usepackage{amsfonts}
\usepackage{color}
\def\newblock{\hskip .11em plus .33em minus .07em}

\newcommand{\be}{\begin{equation}}
\newcommand{\ee}{\end{equation}}
\newcommand{\bea}{\begin{eqnarray}}
\newcommand{\eea}{\end{eqnarray}}
\begin{document}
\title{From Over-charging to Like-charge Attraction in the Weak Coupling Regime}
\author{Xiangjun Xing $^{(a)}$$^{(c)}$, 
Zhenli Xu $^{(b)}$$^{(c)}$, 
Hongru Ma$^{(a)}$}
\affiliation{%
$^{(a)}$ 
Department of Physics,  Shanghai Jiao Tong University, Shanghai, 200240 China \\
$^{(b)}$
Department of Mathematics,  Shanghai Jiao Tong University, Shanghai, 200240 China \\
$^{(c)}$
Institute of Natural Sciences, Shanghai Jiao Tong University, Shanghai, 200240 China
}

\date{\today} 
\begin{abstract} 
Despite decades of intensive studies, the effective interactions between strongly charged colloids still remain elusive.   Here we show that a strongly charged  surface with a layer of condensed counter-ions behaves effectively as a conductor, due to the mobile nature of the condensed ions.   An external source charge in its vicinity is therefore attracted towards the surface, due to the image charge effect.  This mechanism leads to correlational energies for counter-ions condensed on two distinct surfaces, as well as for free ions in the bulk.  Generalizing Debye-Huckel theory and image charge methods, we analytically calculate these correlation energies for the two-plates problem, at the iso-electric point, where condensed counterions precisely balance the bare surface charges.  At this point, the effective interaction between two plates is always attractive at small separation and repulsive at large separation.  

\end{abstract}
\maketitle
The effective interaction between charged objects inside an electrolyte or plasma is a problem of fundamental importance to many branches of physical sciences. In past few decades, there has been accumulating experimental and numerical evidences \cite{Tata-PRL-1992,Ito:1994zr,1997Natur.385..230L,Crocker-Grier-PRL-1996,Gelbart:2000fk,Angelini:2003fk,Butler:2003uq,Linse:1999kx,tata-boundpairs-2008} of effective attraction between likely charged objects.  This is in direct contradiction with the classical DLVO theory \cite{Verwey-Overbeek,Derjaguin-Landau}, which always predict repulsion between two likely charged objects.  Among these experimental works, the recent work by Tata {\it et. al.} \cite{tata-boundpairs-2008} is particularly striking: These authors directly imaged bound states formed by clusters of likely charged colloids in highly de-ionized water, with influences of boundaries and multi-valence counter-ions carefully excluded.  The attraction appears to exist at scale of Debye length, which is several hundreds nanometers.  



Multiple mechanisms have been proposed to explain like charge attractions.  Most of them are based on the idea of counter-ion condensation, first discovered by Manning \cite{Manning:1969vn} in highly charged lines.   Shklovskii {\it et. al.} \cite{Shklovskii:1999vn, RevModPhys.74.329} have shown that in the so-called strong coupling limit, where the Bjerrum length is much larger than the average spacing between condensed counter-ions, the latter form a strongly correlated 2D liquid, which locally resembles a Wigner crystal.   Furthermore, the total charge of condensed ions may overtake the bare surface charges so that the effective charge density of a strongly charged surface may be reversed.  This phenomenon is popularly called charge inversion or over-charging, and has been confirmed in numerical simulations.  Furthermore, previous studies  \cite{doi:10.1021/jp960458g,Jho:2011ve,Moreira:2001uq,Scaronamaj:2011uq} seem to indicate that in the strong coupling regime, correlations between densely packed counter-ions induce short range attraction between two charged plates, with a characteristic scale of the Gouy-Chapman length, or inter-ion distance.  





Colloidal systems \cite{1997Natur.385..230L,Crocker-Grier-PRL-1996,tata-boundpairs-2008} however are almost always in the opposite, weak coupling regime, and attractions were seen at much larger scale.  Like-charge attraction in colloidal systems therefore can not be explained by the previous theories, and remains a fundamental challenge.   

The shear existence of like-charge attraction indicates that the colloidal systems must be close to the threshold of charge inversion.  The mechanism of charge inversion in the weak coupling regime is presumably very different from that in the strong coupling limit, and may very well be of chemical nature.     
In the spirit of pragmatism, we shall assume charge inversion does happen in the weakly charged regime, and explore its consequences.  The counter-ions close to the surface then form a dilute 2d gas.  We expect that a fraction of these ions can diffuse in the lateral directions.  The free ions in the bulk, on the other hand, forms a dilute 3d plasma, and will be treated using standard Poisson-Boltzmann theory.  The basic idea of this two-fluid model has been explored previously by different groups \cite{Levin:1999ys,Lau:2002fk}, in various depths. 
 By studying the responses of the 2d plasma to external perturbations, we can calculate the correlation energy of each ion due to its proximity to two distinct colloidal surfaces, be it condensed on the surfaces, or freely diffusing in the bulk.  This allows us to obtain analytic result about the correlation-induced effective interaction between two charged plates at the iso-electric point.  

The combination of a (negatively) charged colloidal surface and a layer of condensed counter-ions shall be called a {\em dressed surface}.  The standard electrostatic boundary condition is 
\be
\left. \epsilon' \frac{\partial \Phi'}{\partial n} 
\right|_{\rm surface}
- \left. \epsilon \frac{\partial \Phi}{\partial n} 
\right|_{\rm surface}
= \sigma_{\rm net}(x), 
\label{BC-simga_net}
\ee
where $\Phi,\Phi'$ are the potentials inside the electrolyte and inside the colloid, respectively, both evaluated near the surface.   The dielectric constant of the colloid $\epsilon'$ is typically much smaller than that of water $\epsilon \approx 80$.  It is therefore reasonable to take the limit $\epsilon'/\epsilon \rightarrow 0$, so that $\Phi'$ drops out of the problem altogether.  This approximation however does not change our essential result.  



The net surface charge density $\sigma_{\rm net}(x)$ in Eq.~(\ref{BC-simga_net}) includes both some fixed surface charges density and some mobile condensed counter-ions $q n_c(x)$.  At the level of mean field theory, the latter is related to the local potential via the Boltzmann distribution:
$\sigma_c(x) = q\, n_c \exp{ - \beta q \Phi(x)}$.   Assuming weak perturbation, we may linearize this relation.   At the so-called ``iso-electric point'', the bare surface charges are completely neutralized by the condensed counter-ions, so that the dressed plate is overall charge neutral.  The zero-th order term in $\sigma_c(x)$ is then precisely cancelled by the fixed surface charge density.  Therefore Eq.~(\ref{BC-simga_net}) is linearized into the following homogeneous {\em Robin} boundary condition: 
\be
{\mu}\frac{\partial \Phi}{\partial n} =   \Phi(x),
\quad \quad 
\mu = \frac{\epsilon}{n_c \beta q^2} . 
\label{BC-composite-1}
\ee
If all condensed counter-ions were mobile in the transverse directions, 
the length scale $\mu$ would be just the Gouy-Chapman length, and takes the value of a few $\AA$ to a few nm for strongly charged surfaces and therefore is much smaller than sizes of colloids and separation between them. 
It is  therefore attempting to ignore the left hand side of Eq.~(\ref{BC-composite-1}) altogether.  The boundary condition then reduces to that for a {\em conductor}.  The idea that a charged surface with condensed counter-ions behaves as a conductor was first proposed by Shklovskii {\it et. al. } \cite{RevModPhys.74.329,Shklovskii-JCP-2000}, in the context of charge inversion theory.   
It shall play important role in our analysis of counter-ion correlations below. 

To see how a dressed surface behaves as a conductor, consider a dielectric sphere (with radius $a$ and $\epsilon' \approx 0$) at the iso-electric point inside a uniform external field $E_0$ (no electrolyte).  The potential outside the sphere can be expanded in terms of spherical harmonics:
\be
\Phi(x)  =  - E_0 r \cos \theta + \frac{p \cos \theta}{4 \pi \epsilon r^2}, 
\label{Phi-out}
\ee
where $p$ is the induced dipole moment of the dressed sphere.   Substituting this back into Eq.~(\ref{BC-composite-1}) we find:
\be
p = 4 \pi a^3 \epsilon E_0 
\left( \frac{\epsilon_{\rm eff}/\epsilon -1}{\epsilon_{\rm eff}/\epsilon + 2} \right), 
\quad \epsilon_{\rm eff} = \epsilon \frac{a}{\mu}. 
\label{e-eff-sphere}
\ee
For a typical colloid $a \sim 100 nm \gg \mu \sim \AA$, hence the effective dielectric constant $\epsilon_{\rm eff}$ of the dressed sphere is much larger than that of water. The sphere indeed behaves like a conductor.  

To calculate the correlation energy for an ion, we shall follow Debye-Huckel's strategy.   First consider an external point charge $Q$ at a distance $d$ from a  dressed plate at the iso-electric point.  As long as $d \gg \mu$, we can again ignore LHS of Eq.~(\ref{BC-composite-1}) and treat the plate as a conductor.  Within linearized PB theory, then, the plate creates an opposite {\em point} image charge in the other side which exerts a reaction potential 
\be
\phi_Q = - \frac{1}{4 \pi \epsilon} \frac{Q}{2 d} e^{- 2 \kappa d}
\ee
on the source ion, and attracts it towards the plate. 

\begin{figure*}
\begin{center}
\includegraphics[width=16cm]{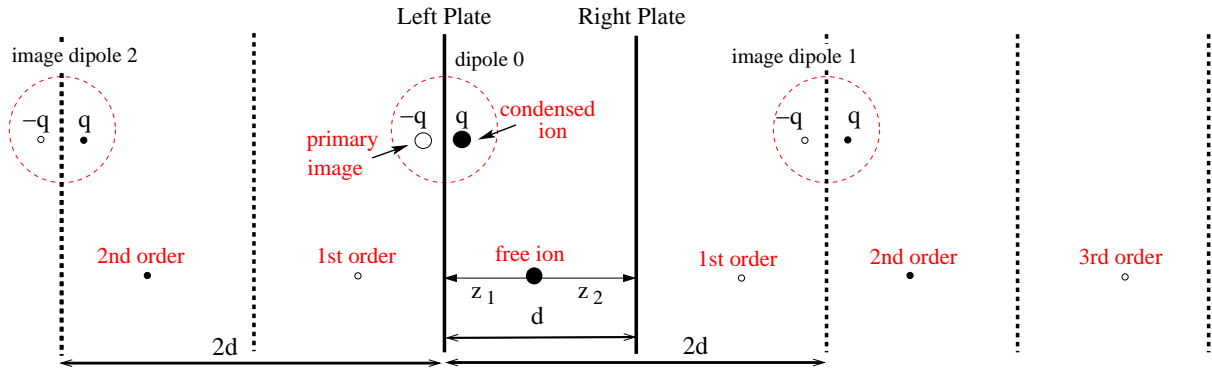}
\caption{ Two parellel plates, shown as vertical solid lines, are separated by distance $d$.  A condensed ion $q$ on the left plate is shown as a big black dot. (It is drawn away from the plate to make it more visible. )  It has a primary image $-q$ (big open dot) just to the left of the left plate.  The source charge and the primary image form the dipole 0, with a dipole moment $p$, pointing to the right.  The right plate further creates an image dipole (dipole 1) with the same magnitude and orientation, at a distance $2d$ from dipole 0. The left plate  further produces another image dipole  (dipole 2), at a distance $2d$ to the left of the left plate.   These two image dipoles are responsible for the lowest order reaction potential acting on the condensed ion $q$.   The effects of higher order images dipoles are smaller by powers of $e^{- 2 \kappa d}$.  Also shown in the lower part are a free ion between two plates, as well as a few low order images.  All images interact with source ion.    }
\label{fig:two-plates}
\vspace{-5mm}
\end{center}
\end{figure*}

What if the source charge $Q$ is one of the counter-ion condensed on the plate?  In this case, the left hand side Eq.~(\ref{BC-composite-1}) can no longer be neglected.  More importantly, the approximation of 2d fluid for condensed counter-ions is no longer valid at scale shorter than the average separation between ions.  Nevertheless, far away from the plate, the dressed plate should still behaves as a conductor.  Its effect can be described by a point image $-q$,  called primary image,  shown in Fig.~\ref{fig:two-plates}.  The combination of source and images behaves effectively as a dipole, called ``dipole zero''.   {In the regime of low density $n_c$, a simple calculation shows that the dipole moment $p$ is given by $p = 2 q \mu$.}  If the density $n_c$ is not low, however, the dipole moment $p$ is rather difficult to calculate, and shall be reserved for a separate publication.  At this stage, we shall treat $p$ as a fitting parameter.  

Now introduce another identical dressed plates, i.e. the right plate in Fig.~\ref{fig:two-plates}, at a distance $d$ from the left plate.   Within the linearized Poisson-Boltzmann theory, the effect of the right plate is to introduce more image dipoles.  To the lowest order, an image dipole at distance $2d$ to the right of the source arises, which shall be referred to as {\em image dipole 1}, see Fig.~\ref{fig:two-plates}.  It has the same orientation and moment as the dipole 0.  Image dipole 1 is reflected again by the left plate, and gives rise to the {\em image dipole 2}, at a distance $2d$ to the left of the left plate. 

\begin{figure}[h]
\begin{center}
\includegraphics[width=7cm]{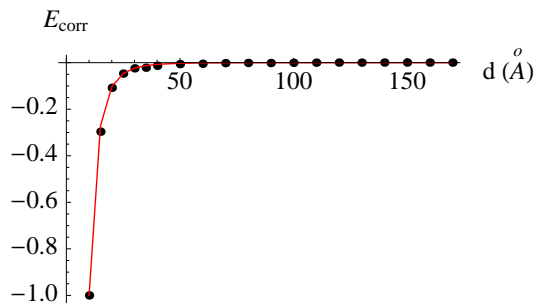}
\caption{Correlation energy of condensed counter-ions as a function of interplate distance.  Dots: simulation of the implicit two-fluid model; line: analytic prediction Eq.~(\ref{U_c-tot}).  The data are renormalized to be unity at the shortest separation $d = 10 \AA$.  In this simulation, there are 40 point-like monovalence counter-ions condensed on to each plate with size $100\AA\times 100\AA$.  The boundary conditions are taken into account using image charge methods.  The system is deep in the weak coupling regime. }
\label{Ecorr-simu}
\end{center}
\vspace{-5mm}
\end{figure}

Dipole 1 and dipole 2 are responsible for the leading order correlation potential acting on the source charge $q$.  Moreover, the interaction between the image dipole 2 and the source charge $q$ is identical to that between the image dipole 1 and the primary image.  It  follows that the lowest order correlation energy of the source charge is  half \footnote{The extra $1/2$ is common to all the image charge interactions.}  of the interaction energy between dipole 0 and the image dipole 1, which, in the linearized Poisson-Boltzmann theory, is given by 
\bea
u_c &=& \left. - 
\frac{1}{2} \cdot \frac{1}{4 \pi \epsilon} 
(p \partial_z)^2 \left( {z}^{-1}{e^{-\kappa z}} \right)
\right|_{z = 2 d}
\nonumber\\
&=& - \frac{p^2}{8 \pi \epsilon} 
\frac{1}{ 4 d^3} 
\left(  1 + 2 \kappa d + 2 (\kappa d)^2
\right)
e^{- 2 \kappa d}.
\label{u-single-charge}
\eea
As expected, this correlation energy is always negative.   Since all counter-ions are identical, their total correlation energy is simply Eq.~(\ref{u-single-charge}) multiplied by the total number of counter-ions on two plates.   Furthermore it is rather straightforward to include all higher order image dipoles.  The total correlation energy of condensed ions is:
\bea
U_c =  & - & \frac{A n_c p^2 }{16 \pi \epsilon d^3}
  \left[  \text{Li}_3(e^{-2 {\kappa d}}) 
+ 2 \kappa d\, \text{Li}_2 (e^{-2 { \kappa d}})
\right.  \nonumber\\
& & \quad \quad \quad - \left.
2 (\kappa d)^2 \log\left(1 - e^{- 2 {\kappa d}}\right)
\right],
\label{U_c-tot}
\eea
where $\text{Li}_m = \sum_{k=1}^{\infty} z^k/k^m$ are polylog functions, and $A$ is the total area of the plates.  This result is in excellent agreement with direct Monte Carlo simulation of the implicit two fluids model, where the free ions between two plates are treated using linearized Poisson-Boltzmann theory, as shown in Fig.~\ref{Ecorr-simu}.  This clearly establishes the image induced dipole-dipole interaction as a main mechanism of electrostatic correlation in strongly charged colloidal systems.  

The same method can be used to calculate the correlation energy of the free ions between two plates.   A free ion (big closed circle in the lower part of Fig.~\ref{fig:two-plates}) between two plates  has an infinite sequence of image charges in its each side.  A few low order images are shown in Fig.~\ref{fig:two-plates}.   Since the dressed plates behave as conductors, all odd order images (small open circles) have opposite charges $-q$, and are at distances $2z_{1,2},  2(z_{1,2} + d), 2(z_{1,2} + 2d), \ldots $ from the source charge, where $z_{1,2}$ are the distances from  the source to the left/right plate respectively.   All even order images (small closed circles) have like charges $q$, and are at distances $2 d, 2 \cdot 2d, 2 \cdot 3d, \ldots$ from the source charge.   Summing up the interaction between the source charge and all its images, and integrating over the space between plates, we obtain the total correlation energy for all free ions between plates as \footnote{Note that we assume that the correlation energy for each ion is sufficiently small that it does not change the uniform density distribution of free ions between plates.} 
\bea
2\, U_{f} &=& - \frac{q^2A2 n_0}{4 \pi \epsilon}  \int_0^d dz_1 \sum_{l = 0}^{\infty} \left( 
\frac{e^ { - 2 \kappa (z_1 + ld)}}{2(z_1 + ld)}
+ \frac{e^ { - 2 \kappa (z_2 + ld)}}{2(z_1 + ld)}
\right) \nonumber\\
&+ &   \frac{q^2 2\,A\, 2 n_0}{4 \pi \epsilon}  \int_0^d dz_1 
 \sum_{l = 1}^{\infty}
 \frac{e^ { - 2 \kappa l d }}{2 l d }.  
 \label{U_f_total}
\eea
By simple variable transformations $z_{1,2} + l d \rightarrow z$, the first integral (due to odd order images) can be transformed into
$\int_0^{\infty } dz\, z^{-1}{e^ { - 2 \kappa z}}$, 
which is independent of $d$, and therefore does not contribute to the effective interaction between plates.  The apparent logarithmic divergence at small $z$ is due to our improper treatment of short scale details, but has no effect on our result.  The second integral in Eq.~(\ref{U_f_total}) (due to even order images) can be easily calculated: 
\be
U_f = -  \frac{q^2 A n_0}{4 \pi \epsilon} 
\log\left( 1 - e^{- 2 {\kappa d}} \right),
\label{U_f_final}
\ee
which, somewhat surprisingly, is always positive.   Numerical simulation of this correlation energy shall be carried out in a forthcoming publication.  

\begin{figure}[thb!]
\begin{center}
\includegraphics[width=7cm]{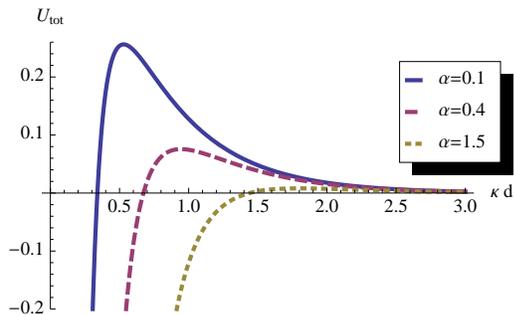}
\caption{The total correlation energy for three cases.  The unit of $U_{\rm tot}$ is $q^2 n_0 A /4 \pi \epsilon$.  }
\label{fig:U_tot}
\end{center}
\vspace{-5mm}
\end{figure}

The total correlation energy for the two-plates system is the sum of Eq.~(\ref{U_c-tot}) and Eq.~(\ref{U_f_final}).  At the level of our approximation, it is also the interacting free energy between two charged plates.   In the large separation regime $\kappa d \gg 1$,  $U_c \sim - e^{- 2 \kappa d}/\kappa d$, while $U_f \sim e^{- 2 \kappa d}$.  Therefore $U_f$ always dominates and two plates repel each other.  By contrast, in the small separation regime $\kappa d \ll 1$, $U_c \sim - 1/\kappa d^3$, while $U_f \sim -\log(\kappa d)$.  Hence $U_c$ always dominates and two plates attract.  The quantitative behaviors of $U_{\rm tot}$ is controlled by one dimensionless parameter $\alpha = n_c p^2 \kappa^3/n_0 q^2$.  Three representative cases of various values of $\alpha$ are shown in Fig.~\ref{fig:U_tot}.  

The result Eq.~(\ref{U_f_final})  describes the effective interaction between two neutral conductor plates.  Interestingly enough, it also describes the interaction between two neutral dielectric plates with $\epsilon' \ll \epsilon \approx 80$.  In the latter case, all images are likely charged, and the total correlation energy is still given by Eq.~(\ref{U_f_total}), but with the first integral changing sign.  This integral is however independent of $d$ as we have shown above.  Therefore the total correlation energy is still given by Eq.~(\ref{U_f_final}).  

We emphasize that the essential ingredient of our analysis is the fact that mobile condensed counter-ions make the charged surfaces resemble conductors.  This is clearly independent of the dielectric constant of the charged colloids themselves.  Our approach therefore applies to arbitrary strongly charged objects in arbitrary solvent.  

In the regime of low density of mobile counter-ions, we have $p\sim 2 q \mu \propto 1/n_c$, hence the prefacton of Eq.~(\ref{U_c-tot}) is inversely proportional to mobile counter-ion density $n_c$.  \footnote{This relation of course does not hold for arbitrary small $n_c$.  Because if the length scale $\mu$ in Eq.~(\ref{BC-composite-1}) becomes comparable with the separation $d$, our dipole approximation of the image charges is no longer valid. } Rather ironically, therefore, the correlation induced attraction is stronger when there are fewer mobile counter-ions.   The surface charge density on a real surface should be heterogeneous at atomic scales.   This may lead to pinning of most counter-ions at the threshold of iso-electric point, and therefore may greatly enhance correlation-induced attraction between two such charged colloids.  

We suspect that strongly charged colloids in highly deionized water may be very close to the iso-electric point, due to some unknown self-tuning mechanism, that is probably of chemical nature.  It this is so, the correlational energies Eq.~(\ref{U_c-tot}) and Eq.~(\ref{U_f_final}) are still valid.  Direct Coulomb interaction between two charged plates proportional to $\sigma^R_1 \sigma^R_2$ should also be included, where $\sigma^R_{1,2}$ are the renormalized surface charge density.  Depending on the relative signs of $\sigma^R_{1,2}$, this could be either repulsive or attractive.   For the case of colloids with finite sizes, additional attractions arise due to the overall polarization of colloids by nearby charged colloids, whose magnitudes are proportional to $(\sigma_{1,2}^R)^2$, and is also doubly screened.   This effect was analyzed by Levin \cite{Levin:1999ys}, as well as by Zhang and Shklovskii \cite{Zhang:2005kx}.  The quasi-conductor nature of the dressed surface also played an essential role in their analyses as well.  We note that in the proximity of the iso-electric point, all these effects are subdominant to the effective interaction derived in our work.  


Detailed understanding of short scale physics and chemistry is necessary in order to determine the values of  parameters used in our two-fluids model, such as the effective/renormalized charge density of a highly charged surface, as well as the effective density of mobile condensed counter-ions.  This demands close collaborations between theorists and experimentalists, as well as between physicists and chemists.  

X. X. thanks Shanghai Jiao Tong University for financial support, and thanks Erik Luijten, Michael Brenner, Anatoly Kolomeisky, Penger Tong,  Leo Radzhiovsky, and J.-F. Joanny for stimulating discussions on electrolytes and colloidal physics over the years.  Z. X.'s research is supported by the Chinese Ministry of Education (NCET-09-0556) and NSFC (No.11026057). H. M. acknowledges financial support by the National Natural  Science Foundation of China (No.10874111).



\end{document}